\documentclass[11pt,twoside]{article}
\usepackage{./asp2014}

\aspSuppressVolSlug
\resetcounters

\begin{document}

\title{Circumbinary Components of Contact Binaries}
\author{Osman Demircan,$^1$ and \.{I}brahim Bulut$^1$
\affil{$^1$Department of Space Sciences and Technologies, Faculty of Arts and Sciences, \c{C}anakkale Onsekiz Mart University, \c{C}anakkale, Turkey; \email{demircan@comu.edu.tr}}}

\paperauthor{Osman Demircan}{demircan@comu.edu.tr}{ORCID_Or_Blank}{Author1 Institution}{Department of Space Sciences and Technologies, Faculty of Arts and Sciences, \c{C}anakkale Onsekiz Mart University}{\c{C}anakkale}{State/Province}{TR-17020}{Turkey}
\paperauthor{\.{I}brahim Bulut}{ibulut@comu.edu.tr}{ORCID_Or_Blank}{Author2 Institution}{Astrophysics Research Centre and Observatory, \c{C}anakkale Onsekiz Mart University}{\c{C}anakkale}{State/Province}{TR-17020}{Turkey}

\begin{abstract}
The period changes of contact binaries obtained by the analysis of eclipse minima timing are found mostly chaotic in nature. However, they are representable by a few cyclic changes superposed on a secular  change. The cyclic changes are caused most probably by the third components revolving around the contact binaries. Some typical examples  of the period changes of contact binaries  are presented in the present contribution. 

\end{abstract}

\section{Overview, Ongoing Work and Some Results}

The incidence of third components in binary systems increase with decreasing orbital period $P$, from 50\% at $P$ = 9 d to probably 100\% at $P$ = 1 d \citep{d08}. Thus, all contact binary stars may exist in multiple systems.

Sinusoidal modulation of the orbital period variations of binary stars are caused mostly by the light time effect (LITE) of unseen third components.  The LITE interpretation for the cyclic period changes of contact binaries was introduced as early as 1990's  \citep[see e.g.][]{d01, d04}. It was extended later to many contact binaries TY Boo, TZ Boo, CK Boo, GW Cep, UX Eri, V566 Oph and BB Peg. 

The period changes of some well known contact binaries with largest number of available minima times were studied again by reconsidering the possible LITE effects in their \textit{O-C} diagrams formed by the most accurate minima times. The preliminary results on the first example W UMa is given in Fig. 1 and Table 1. 

As seen in Fig. 1 and Table 1, accurate eclipse timing of the low mass contact binaries may reveal the evidence on long period low mass companions, even for distant Jovian planets of contact binaries.


\begin{table}
\caption{Parameters for the LITE Orbits of W UMa.}
\smallskip
\begin{center}
{\small
 \begin{tabular}{lcccc}
 \tableline
 \noalign{\smallskip}
 Parameter&Unit&T3&T4&T5\\
 \noalign{\smallskip}
\tableline
\noalign{\smallskip}

$T_{0}$ &HJD&&2435918.4229(13)&\\
$P$ &day&&0.33363661(4)&\\
$Q$&day&&$-2.4(5) \times 10^{-7}$&\\
$P_{3,4,5}$&yr&95(3)&17.91(21)&13.61(23)\\
$T_{ per3,4,5}$ &HJD&2410271(1905)&2457143(506)&2458503(568)\\
$A$ &day&0.02902(38)&0.0019(11)&0.00094(11)\\
$\dot{\omega}$&deg&87(11)&0(27)&0(36)\\
$e$&&0.092(27)&0.21(12)&0.28(21)\\
$f$ ($m_{3,4,5}$)&M$_{\odot}$&0.01389&0.00011&0.000027\\
$m_{min3,4,5}$&M$_{\odot}$&0.3890&0.0806&0.0439\\

\noalign{\smallskip}
\tableline\
\end{tabular}
}
\end{center}
\begin{small}

Note: $P_{3,4,5}$, $A$, $f$ ($m_{3,4,5}$) and $m_{3,4,5}$ ($i$=90 deg) are the LITE periods, semi-amplitudes, the mass functions and masses. T3, T4 and T5 represents LITEs due to three unseen objects in the system. \\
\end{small}

 \end{table}


\begin{figure}
\begin{center}

{\includegraphics[bb= 25 230 410 800, width=7.2 cm]{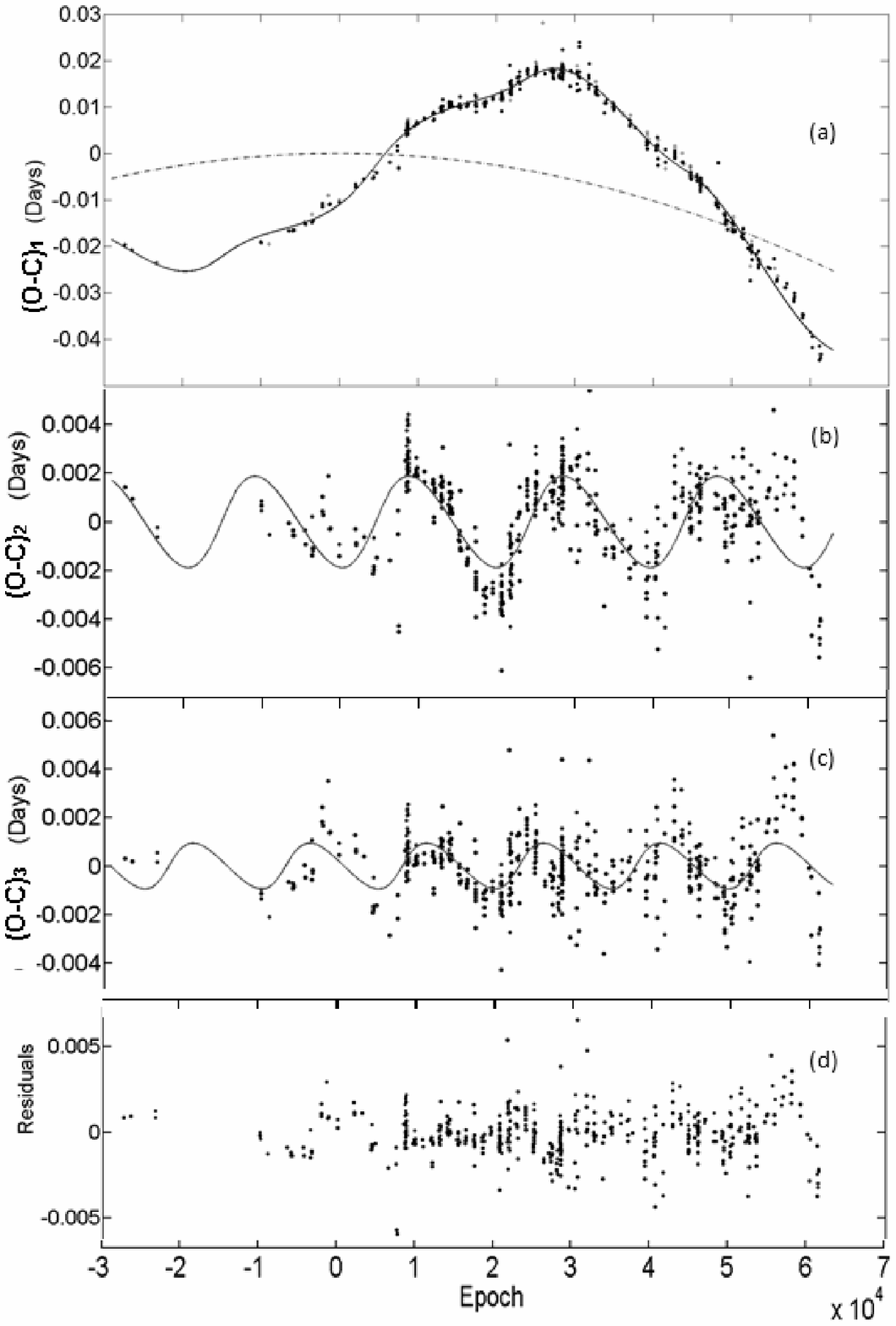}}

\end{center}
\caption{The \textit{O-C} diagram of W UMa represented by a parabola and the sum of two sine curves (a). The residuals from the parabola plus first sine curve (b). The residuals from the best fit to first residuals (c). The third residuals from best fit to second residuals (d). All residuals imply light effects due to additional unseen objects in the system.} \label{fig2}
\end{figure}




\begin{thebibliography}{}

\bibitem[\protect\citeauthoryear{Demircan et al.}{1992}]{d01}Demircan, O., Derman, E., $\&$  Muyessero\u{g}lu, Z. 1992, $A$\&$A$, 263, 165

\bibitem[\protect\citeauthoryear{Demircan}{1994}]{d04}Demircan, O. 1994, in The Impact of Long-Term Monitoring on Variable Star Research: Astrophysics, Instrumentation, Data Handling, Archiving, Proceedings of the NATO Advanced Research Workshop, held in Ghent, Belgium, November 15-18, 1993, Dordrecht: Kluwer, 1994, edited by Christiaan Sterken and Mart De Groot. NATO ASI Seri C, 436, 77

\bibitem[\protect\citeauthoryear{Tokovinin et al.}{2006}]{d08}Tokovinin, A., Thomas, S., Sterzik, M., $\&$ Udry, S. 2006, $A$\&$A$, 450, 681

\end{thebibliography}
\end{document}